\documentclass[onecolumn,preprintnumbers,amsmath,amssymb,nofootinbib,showkeys]{revtex4}

\usepackage{graphicx}
\usepackage{dcolumn}
\usepackage{amsmath}          
\usepackage{amssymb}
\usepackage{bm}
\usepackage[T1]{fontenc}
\usepackage[latin9]{inputenc}
\usepackage{esint}
\usepackage[brazil,english]{babel}
\usepackage[usenames,dvipsnames]{color}
\usepackage{longtable}
\usepackage{ulem}
\usepackage{nicefrac}

\begin{document}

\title{Dynamical model for primordial black holes}

\author{F. Ruiz}
\email{fruiz@usp.br}
\affiliation{Universidade de S\~{a}o Paulo, Instituto de F\'{\i}sica,
 Caixa Postal 66318, 05315-970, S\~{a}o Paulo-SP, Brazil}

\author{C. Molina}
\email{cmolina@usp.br}
\affiliation{Universidade de S\~{a}o Paulo, Escola de Artes, Ci\^{e}ncias e Humanidades, Avenida Arlindo Bettio 1000, CEP 03828-000, S\~{a}o Paulo-SP, Brazil}

\author{J. A. S. Lima}
\email{jas.lima@iag.usp.br}
\affiliation{Universidade de S\~{a}o Paulo, Instituto de Astronomia, Geof\'{\i}sica e Ci\^{e}ncias Atmosf\'{e}ricas, Rua do Mat\~{a}o 1226, CEP 05508-090, S\~{a}o Paulo-SP, Brazil}

\begin{abstract}

 Primordial black holes are analytically and numerically discussed based on the extended McVittie spacetime solution. By assuming that dark matter and radiation are the only sources of energy accreted by the forming central object, it is found that the black-hole mass evolution depends on the initial mass of the seed, the time in which the black hole emerges, and also on the average peculiar velocity of dark matter particles. Constraints on the initial conditions of the primordial black holes are derived from profiles of the  black-hole accretion mechanism and cosmological environment. A large range of masses is compatible with our approach. In particular, masses of the order of $10^{10}M_{\odot}$ today may also be generated from small seeds. An incubation time for the emerging horizons is observed when the initial masses of the seeds are close to the particle-horizon mass. It is also argued that the McVittie-type description is consistent with the Schwarzschild solution as long as other astrophysical processes near the central object are neglected. 

\end{abstract}


\maketitle

\section{Introduction} 

The possible formation and evolution of compact objects in the primeval Universe has been extensively studied and reformulated in the last decades. Long ago Ambartsumian suggested that stellar clusters and galaxies would be formed from the expansion of dense objects named protostars \cite{RevModPhys.30.944,1960SvA.....4..187A}. Novikov's calculations also showed that fluctuations on the metric as well as on the primordial density field would generate inhomogeneities in scales of the order of $100Mpc$ today \cite{Novikov1}. In addition, by assuming a closed Friedmann-Lema\^{\i}tre-Robertson-Walker (FLRW) spacetime region after the initial singularity, which initially did not expand with the rest of the Universe (considered a flat FLRW geometry), an exact solution of Einstein's equations was also derived \cite{1965SvA.....8..857N}. 
Many  versions of Novikov's model  were also suggested. In general, the compact inner region is described by a perturbed Friedmann metric smoothly connected with the FLRW flat model using a transition region (see, for instance \cite{1979A&A....80..104N}). By that time, the term ``primordial black hole'' (PBH) was already in use.

The detection of gravitational waves by the LIGO/VIRGO Collaboration \cite{A2016a} and the first image of a black hole in the nearby radio galaxy M87 recently obtained by the Event Horizon Telescope \cite{Akiyama:2019bqs} opened a remarkable observational window for the understanding of the Universe and the possible existence of PBHs. As suggested by Nakamura, Thorne \textit{et al.} \cite{Nakamura_1997}, or more recently by Raidal \textit{et al.} \cite{Raidal_2017}, PBH mergers would be relevant sources of gravitational waves. In fact, the most probable scenario to explain the LIGO detection GW150914 \cite{A2016a} corresponds to the coalescence of a PBH binary system \cite{PhysRevLett.117.061101}. 

Moreover, the hypothesis that PBHs may account for either a part or even the whole dark matter is still disputed, with recent proposals for PBH mass spectra which would solve this and other problems \cite{carr-silk}. In particular, PBHs with masses of the order of Earth mass, $M_{\textrm{PBH}} \approx M_{\oplus}(125 GeV/T^{2})^{2}$, may also be generated in the radiation phase during a first-order phase transition in the vicinity of the electroweak scale \cite{PRD97}.  More recently, it was also suggested that the excess in microlensing events in the 5-year Optical Gravitational Lensing Experiment data together with the anomalous orbits trans-Newtonian objects can be interpreted as a new population of dark objects (presumed to be PBHs) captured by the solar system instead of the free-floating Planet 9 hypothesis \cite{PhysRevLett.125.051103}.

The timescale for the evolution of PBHs that could exist until today is roughly of the order of $1/H_0$ (where $H_0$ is the Hubble constant) since they are supposed to be formed in the radiation dominated era \cite{PhysRevD.94.083504,PhysRevD.59.124013}. In principle, such objects leave traces that eventually can be observed and used for identifying them in different mass scales. For instance, promising observational windows involve gravitational lensing, interaction with the baryonic matter and the emitted radiation by the infalling matter during the accretion process \cite{Sasaki_2018}. Within this timescale, accretion processes could play an important role in the spacetime geometry. Hawking and Carr \cite{10.1093/mnras/168.2.399} developed a similarity solution in which a PBH grows at the same rate as the particle horizon but it fails to satisfy the boundary condition expected physically. Accretion dynamics were also treated as a quasistationary process in \cite{1967SvA....10..602Z,Shapiro:1983du,NAYAK2011}. In particular, accretion of dark matter and dark energy were considered in \cite{LIMA2010218}.

There are exact solutions of Einstein's equations which can be interpreted as spacetimes supported by a compact object immersed in a cosmological background. Some of them are the solutions first obtained by Kottler \cite{SdSK} (also known as Schwarzschild-de Sitter/anti--de Sitter), McVittie \cite{McVittie:1933zz} and Thakurta \cite{Thakurta}. Several other candidates describing cosmological black holes were also discussed in the literature \cite{Faraoni:2007es,Faraoni:2008tx,Firouzjaee:2008gs,Firouzjaee:2010ia,Moradi:2015caa,Nolan:1998xs,Nolan:1999kk,Nolan:1999wf}. In particular, the McVittie spacetime assumes the existence of a central and spherical inhomogeneity immersed in a perfect fluid. It has an asymptotically FLRW behavior and a spacelike big bang singularity, and, as such, it lacks a global timelike Killing vector field. The physical interpretation of the McVittie solution has been widely debated since it was proposed and some controversy  has also been generated \cite{Nolan:1998xs,Nolan:1999kk,Nolan:1999wf,Faraoni:2008tx,Lake:2011ni}. Nevertheless, it was shown that in a cosmological Einstein-de Sitter background, the McVittie solution is reduced to the Schwarzschild-de Sitter metric, and the central object effectively describes a black hole.

The Einstein's equations for the McVittie geometries can be solved by specifying two functions $a(t)$ and $\mu(t)$ and it should be given some information of the energy-momentum tensor as suggested in \cite{carrera}. The original McVittie's work has $\mu(t)=m/a(t)$ where $m$ is a constant, but it was extended and the properties of the generalized McVittie metric were explored by several authors \cite{carrera,Guariento:2012ri,daSilva:2012nh,Faraoni:2014nba,daSilva:2015mja}. The general McVittie-type solution \cite{carrera,Guariento:2012ri} describes a compact object with a time-varying mass ($dm/dt \neq 0$) so that the associated spacetime can be adopted to model the accretion of mass by a black hole. 

In the present work, primordial black holes are described through an extended McVittie spacetime, by imposing a simple accretion mechanism driving the black-hole mass in a FLRW background along the lines developed in \cite{1967SvA....10..602Z,Shapiro:1983du,NAYAK2011}. Our main goal is to examine a dynamical model for the PBH evolution immersed in a cosmological environment. The generalized McVittie metric is used as an interpolation tool, linking local accretion physics with the large-scale cosmology. The black-hole evolution is obtained through the behavior of its apparent horizons. The large scale cosmological environment is characterized here by the scale factor of the FLRW metric far from the the central inhomogeneity. The short-scale accretion dynamics is fully dependent on the mass function of the central object $m(t)$. As we shall see, the different scales of the PBH system are interpolated by the adopted McVittie-type metric.

The paper is organized as follows. In section~\ref{macvittie-geometry}, the  general characteristics of the generalized McVittie spacetimes are reviewed.  By assuming  suitable mass and scale functions, the specific model for the primordial black holes is proposed in section~\ref{modeling-pbh}. The  model is analyzed with more detail in section~\ref{analizing}, where the relevant parameter space is constrained and the possible horizon dynamics is discussed. The paper is closed with the final remarks in section~\ref{final-remarks}.

\section{Extended M\MakeLowercase{c}Vittie spacetime}
\label{macvittie-geometry}

 A local accretion model for primordial black holes in a cosmological setting may be constructed by interpolating the local  geometry (close to the compact object) and the large scale  cosmological description. In the present work, this interpolation will be carried out with the generalized McVittie spacetime. One may assume that the large scale structure of the Universe is well described by the $\Lambda$ cold dark matter standard cosmological model. It is well known that the combination of independent observations including supernovas,  baryon acoustic oscillations,  gravitational lensing, and the angular power spectrum of the cosmic microwave background suggest a spatially flat universe \cite{planck}. 

In this context, our starting point is that PBHs are formed during the radiation dominated era and evolve capturing  radiation and dark matter from their vicinity in virtue of some accretion mechanism.  To model this physical scenario, let us now consider the extended McVittie-type metric \cite{carrera}
\begin{equation}
ds^2 = -\left[\frac{1-\dfrac{m(t)G}{2c^2a(t)r}}{1+\dfrac{m(t)G}{2c^2a(t)r}}\right]^2c^2dt^2 +\left[ 1+\dfrac{m(t)G}{2c^2a(t)r}\right]^4 a(t)^2\left( dr^2 +r^2d\Omega^2 \right)
\, ,
\label{mcvittiegeneralizado}
 \end{equation}
where $d\Omega^2=d\theta^{2} + sin^{2}\theta d\phi^{2}$. It describes a compact object with a time-varying mass $m(t)$ immersed  in a cosmological background characterized by $a(t)$, the cosmic scale function \cite{carrera,Guariento:2012ri}. The $r$ coordinate is the isotropic radius, with $r>m/2a\,$. Note also the existence of a curvature singularity in the limit $r \rightarrow m/2a$, as can be shown calculating the Ricci scalar \cite{CarreraMatteoDomenico}. 

In passing, at the level of the metric description, the time dependence of the black-hole mass could be formally incorporated in the time-varying gravitational constant $G$ as $m(t) \, G \rightarrow m \, G(t)$ and the same happens with a possible time dependence of the speed of light $c$. However, the variation of the fundamental constants like $G(t)$, $c(t)$ (or both) modify the field equations \cite{Lima-94,Magueijo:2003gj, Balcerzak:2016azv, Lima-94}. Such a possibility is not being discussed here.

It should be noticed that the Weyl contribution to the Misner-Sharp energy based on the line element~\eqref{mcvittiegeneralizado} reads
\begin{equation}
 E_w = m(t) c^2 \, ,
\label{misner-sharp}
\end{equation}
thereby showing that the total black-hole energy is also a time-dependent quantity.

It is convenient to express the line element~\eqref{mcvittiegeneralizado} in terms of the areal radius $\hat{r}$. To that end, the coordinate transformation 
\begin{equation}
\label{isotropgeneralizado}
\hat{r}=a(t)r\left[ 1+\frac{m(t)G}{2c^2a(t)r} \right]^{2}
\end{equation}
allows us to write the generalized McVittie metric~\eqref{mcvittiegeneralizado} in the form \cite{Guariento:2012ri}
\begin{equation}\label{mcvittiegeneralizado2}
ds^2 = - R(t,\hat{r})^2 \, c^2dt^2 + \left\{ \frac{d\hat{r}}{R(t,\hat{r})} - \left[ H(t) + M(t) \left( \frac{1}{R(t,\hat{r})} 
- 1 \right) \right] \hat{r} dt \right\}^2 + \hat{r}^2 d\Omega^2 \,.
\end{equation}
The functions $R(t,\hat{r})$, $H(t)$ and $M(t)$ are defined as
\begin{equation}
R(t,\hat{r}) \equiv \sqrt{1 -\frac{2 m(t)G}{c^2\hat{r}}} \,, \,\,
H(t) \equiv \frac{\dot{a}(t)}{a(t)} \,, \,\,
M(t) \equiv \frac{\dot{m}(t)}{m(t)} \, ,
\label{def-RHM}
\end{equation}
where a dot indicates time derivative. Moreover, restrictions on the metric functions must be guaranteed to ensure that the McVittie-type geometry is a physically acceptable model for an accreting PBH in the FLRW environment. In this case, reasonable conditions are \cite{Faraoni:2013aba,daSilva:2015mja}
\begin{eqnarray}
 \label{condi1}   &m(t) > 0 \, ,\quad \forall t > 0 \,;&\\  
   & \displaystyle \lim_{t \to \infty} H(t) \rightarrow H_{\textrm{asym}}, H_{\textrm{asym}}>0 \,;& 
\label{cond-H}\\ 
  \label{condi3}  &M(t) \geqslant 0 \,, \, \forall \, t > 0 \,;& \\
   \label{condi4} &H(t) > 0 \,, \, \forall \, t > 0 \,;& \\
 \label{m0h0}   &\dfrac{G}{c^3}m_0H(t)<1/3\sqrt{3}\,.&
\end{eqnarray} 
In Eq.~\eqref{cond-H}, $H_{\textrm{asym}}$ is a positive constant representing the asymptotic value of the Hubble constant. In Eq.~\eqref{m0h0}, $m_{0}$ is a positive constant denoting the asymptotic mass of the black hole. Such a condition applies in regions in which the black-hole mass does not change ($M = 0$) and it should be satisfied to guarantee the existence of both horizons \cite{Faraoni:2013aba}.  

One way to describe black-hole geometries is through their horizon structure. In the framework of McVittie-type metrics, the proper description of a black-hole horizon is not straightforward because of the spacetime  dynamics \cite{doi:10.1139/p05-063}. In what follows we will characterize black holes through their apparent horizons. Such structures are defined as the closure of a 3-surface which can be foliated by marginal 2-surfaces. These closed marginal surfaces are defined in such a way that the expansion of the surface geodesic vector null fields vanishes \cite{Faraoni:2015ula}. In this case, it should be emphasized that the spacetime does not need to be asymptotically flat.

The expansion of a null vector field $V$ is defined by the following relation:
\begin{equation}  
  \label{trappedV3}\theta_{(V)}= \frac{1}{\sqrt{h}}\mathcal{L}_{V}\sqrt{h}=\frac{1}{\sqrt{h}}\frac{d}{d\lambda_{V}}\sqrt{h}\, ,
  \end{equation}
where $h$ is the determinant of the induced metric on the hypersurface normal to the null vector field $V$, and $\mathcal{L}_V$ is the Lie derivative throughout the field $V$. It can be expressed equivalently in terms of a derivative with respect to the parameter $\lambda_V$ that generates the null geodesic. A future apparent horizon is defined by the following conditions over constant-time slices:   
\begin{equation}
\theta_{(\ell)} = 0\,, \,\, \theta_{(n)} < 0\,,
\end{equation}
with the subscripts $\ell$ and $n$ denoting outgoing and ingoing null vector fields, respectively. 

The apparent horizons are quasilocal structures, implying that their existence can be probed by finite time measurements without previous knowledge of the causal structure of the whole spacetime \cite{Faraoni:2015ula}. Considering the generalized McVittie metric written in the form~\eqref{mcvittiegeneralizado2}, two past apparent horizons appear and can be determined by
\begin{equation}
\left[ H(t)+\frac{2m(t)GM(t)}{c^2\hat{r}R(t,\hat{r})(1+R(t,\hat{r}))} \right]\hat{r}-cR(t,\hat{r})=0 \, .
\end{equation}
It is worth noticing that the Schwarzschild-de Sitter case shows a similar behavior in these coordinates. However, the analytic extension can reveal the future apparent horizons at the same values of the radial coordinate. Though the analytic extension for the generalized McVittie spacetime is not at our disposal we expect a similar structure \cite{daSilva:2015mja}. By using the definition of $R(t,\hat{r})$ presented in Eq.~\eqref{def-RHM} and rearranging terms, a polynomial equation in $\hat{r}$ for each value of $t$ is obtained,
\begin{equation}
\label{polinomio-1}{H\over c^2}\left(H\!-\!2M\right)\hat{r}^4-2\left[{Gm\over c^4} \left( H\!-\!M \right)^2 \!-\! {M\over c} \right]\hat{r}^3 - \left( 4{Gm\over c^3}M \!+\!1 \right)\hat{r}^2+4{Gm\over c^2}\hat{r} -4{G^2m^2\over c^4} = 0 \, .
\end{equation}

In order to characterize the spacetime, the next step is to calculate explicit solutions for the apparent horizons. Hence, it is necessary to set the metric functions $H(t)$ and $m(t)$, fixing the local accretion mechanism and cosmological environment. This issue will be explored in the next section.

\section{Modeling primordial black holes}
\label{modeling-pbh}

 Primordial black holes are expected to be formed in the radiation dominated era. The absorption cross section of dark matter particles and radiation by black holes suggests that such components are the most important to the PBH formation process.  Actually, the photon  accretion rate cannot be neglected because the energy density of radiation is very large at high redshifts. For large values of $r$, the inflowing energy can be calculated for a Schwarzschild black hole \cite{Shapiro:1983du}. We consider this result as an approximation for the dynamical situation and it is expected to be an overestimation of the accretion rate because the expansion of the Universe limits the region in causal contact with the black hole. Far from the black hole, we simplify our analysis of accretion by taking into account two perfect fluids: a dust of noninteracting massive particles with an equation of state $p_{\infty m}=0\,$ representing dark matter, and a thermal bath of massless particles with $p_{\infty r}=c^2\rho_{\infty r}/3\,$ for photons (see also \cite{cosmo-evo-pbh}). 

For short timescales compared with the age of the Universe, the accretion process can be assumed to be stationary. In this way, the phase space distribution function of the massive particles  $f(\boldsymbol{r},\boldsymbol{v},t)$ is a time-independent quantity. Due to the spherical symmetry, it depends only on  the constants of motion of the problem, that is, the magnitude of the angular momentum and the energy of the particles \cite{Shapiro:1983du}.

In the case of a stationary accretion process the unbounded particles whose energy at $r\rightarrow \infty$ is positive far from the black hole provides the  primary contribution to the growth rate of the PBH. In the case of a monoenergetic particle distribution  at $r\rightarrow \infty$,  the solution of the Boltzmann equation is proportional to a Dirac's delta of the particle energy  which peaks for $E_{\infty}$ \cite{Shapiro:1983du}. The accretion rate for photons can be  separately calculated by using the absorption cross section associated with them. The black-hole mass $m(t)$ will be composed by both contributions, and the evolution equation becomes
\begin{equation}\label{rataacrecion}
\dot{m} =\dot{m}_m+\dot{m}_r= \frac{16\pi G^2\rho_{\infty m}}{v_{\infty}c^2} m^2+\frac{27\pi G^2\rho_{\infty r}}{c^3} m^2 \, ,
\end{equation} 
where the quantities $\rho_{\infty m}$ and $\rho_{\infty r}$ are the energy densities at infinity of particles of mass $\mu_0$ and massless particles, respectively.
The quantity $v_{\infty}$ is the mean value of the magnitude of the particles velocities far apart from the black hole and $m$ is the mass of the central object. Those quantities are functions of the time coordinate $t$ which coincides with the cosmological time for large $r$. 

From the observational viewpoint, it is more convenient to express the cosmological time $t$ in terms of a ``Friedmann redshift'' $z$, defined as
\begin{equation}
z(t) =  \frac{1}{a(t)} - 1 \, .
\label{redshift}
\end{equation}
There is a map between $z$ and the coordinate $t$, and hence $z$ can be used as the time parameter. Strictly speaking, the Friedmann redshift and the redshift $Z$ constructed from the generalized McVitte metric do not exactly coincide. The redshift $Z$ depends on both $t$ and $r$ coordinates, and therefore there is no one-to-one relation between $Z$ and $t$ (reflecting the fact that the McVittie spacetime is not homogeneous). However, the Friedmann redshift $z$ approximates the optical redshift if the light emission is not too close to the horizon, and from a practical point of view is more convenient than $t$ to label the cosmological time.

The Einstein's equations for the generalized McVittie metric~\eqref{mcvittiegeneralizado} far from the black hole have the usual Friedmann form, 
\begin{eqnarray}
\label{F1}&H^2=\dfrac{8\pi G}{3}\rho_{\infty} \, ,&\\[3pt]
\label{F2}& \dot{\rho}_{\infty}+3H\left(\rho_{\infty}+\displaystyle {p_{\infty}\over c^2}\right)=0 \, . &
\label{energy-conservation}
\end{eqnarray}
The density and pressure can be decomposed in the usual components: matter, radiation, and cosmological constant (considering a zero-curvature cosmology). Each component has an equation of state, 
\begin{eqnarray}
&\rho_{\infty} = \rho_{\infty r} + \rho_{\infty m} + \rho_{\infty \Lambda}\hspace{2pt}, \, &\\
&p_{\infty}=p_{\infty r} +p_{\infty m}+p_{\infty \Lambda}\hspace{2pt} \, ,
\end{eqnarray}
with
\begin{eqnarray}
&p_{\infty r}=c^2\rho_{\infty r}/3\,\,,&\\[3pt]
&p_{\infty m} = 0\hspace{2pt},&\\[3pt]
&p_{\infty \Lambda} = -c^2\rho_{\infty \Lambda}\hspace{2pt}.&
\end{eqnarray}
Since the fluid components are noninteracting, we are able to solve Eqs.~\eqref{F1} and~\eqref{F2} to obtain the densities and the Hubble function $H$ in the three different epochs in which each fluid dominates. In terms of the Friedmann redshift,
\begin{eqnarray}
\label{densidadradiacion}&\rho_{\infty r} =\dfrac{3H_0^2\Omega_{r0}}{8\pi G}(1+z)^4 \,,&\\[6pt]
\label{densidadmateria}&\rho_{\infty m} =\dfrac{3H_0^2\Omega_{m0}}{8\pi G}(1+z)^3 \,,&\\[6pt]
\label{hubblefunction}& H(z)=\left\{
\begin{array}{cc}
H_0\sqrt{\Omega_{r0}}(1+z)^2 \,\, , \,& z\geq z_{eq}\,,\\[6pt]
H_0\sqrt{\Omega_{m0}}(1+z)^{3/2} \,\, , \, & \bar{z}_{eq}\leq z<z_{eq}\,,\\[6pt]
H_0\sqrt{\Omega_{\Lambda 0}} \,\, , \, & z<\bar{z}_{eq}\,, 
\end{array}
\right.
&
\end{eqnarray} 
where $\Omega_{r0}$, $\Omega_{m0}$ and $\Omega_{\Lambda 0}$ are the density parameters today for radiation, matter, and cosmological constant respectively. The quantities $z_{eq}$ and $\bar{z}_{eq}$ denote the Friedmann redshift for the matter-radiation equality and the matter-cosmological constant equality moments. 

On the other hand, the mean value of the magnitude of the peculiar velocity of particles in the comoving frame in a FLRW geometry should be matched with the mean value of the magnitude of the velocity of the particles at infinity in Eq.~\eqref{rataacrecion}, given by
\begin{equation}
v_{\infty}^k=a\frac{u^k}{u^0}\,,
\end{equation}
where $u^{\nu}$ is the four-velocity of the particle in a comoving frame at the FLRW geometry. As a consequence of the spatial translations invariance, we can write
\begin{equation}
\mu a(t) \frac{v_{\infty}^k(t)}{\sqrt{1-\dfrac{v_{\infty}^2(t)}{c^2}}}=b^k\,,
\label{b-general}
\end{equation}
with $b^k$ being a constant with respect to the time coordinate. Considering the constant $b^k$ today,
\begin{equation}
b^k=\mu \frac{v_{\infty 0}^k}{\sqrt{1-\dfrac{v_{\infty 0}^2}{c^2}}} \, .
\label{b-today}
\end{equation}
Combining Eqs.~\eqref{b-general} and \eqref{b-today}, the magnitude of the peculiar velocity of the particles in terms of the scale factor $a$ is obtained:
\begin{equation}\label{vinfty-a}
v_{\infty}(a)=\frac{v_{\infty 0}}{\sqrt{1-\dfrac{v_{\infty 0}^2}{c^2}}\sqrt{a^2+\dfrac{v_{\infty 0}^2/c^2}{1-v_{\infty 0}^2/c^2}}}\,.
\end{equation}

The velocity of dark-matter particles today is an unknown quantity, but assuming the cold dark matter scenario it is expected to be much smaller than the speed of light. For example, the typical velocity of a galaxy in a cluster of galaxies is  $v_{\infty 0} \approx 10^{-6}pc/yr$, which is a small fraction of $c$. The expected mean velocity we are interested in must be at most of the order of that velocity (we will return to that point in the next section). Then, we can use the result~\eqref{vinfty-a} in its nonrelativistic approximation, that also can be expressed in terms of the Friedmann redshift $z$ as
\begin{equation}\label{velocidad}
v_{\infty}(z)=v_{\infty 0}(1+z)\,.
\end{equation}

The integration of~\eqref{rataacrecion} can be performed replacing the results~\eqref{densidadradiacion}, \eqref{densidadmateria}, \eqref{hubblefunction} and~\eqref{velocidad} in~\eqref{rataacrecion}. We have,%
\footnote{A previous work \cite{LIMA2010218} considered the accretion rate function due to Babichev \cite{PhysRevLett.93.021102} to study accretion of dark matter and dark energy.}
in terms of $z$,
\begin{equation}\label{masa}
m(z)=\left\{
\begin{array}{cc}
\dfrac{m_{I}}{1-\alpha_1\ln\left| \frac{1+z_I}{1+z}  \right|-\beta_1\left[(1+z_{I})^2-(1+z)^2  \right]} \,\,,& z\geq z_{eq}\,,\vspace{0.5cm} \\ 
\dfrac{m_I}{1-\alpha_2\left[ (1+z_{eq})^{1/2}-(1+z)^{1/2}  \right]-\beta_2\left[(1+z_{eq})^{5/2}-(1+z)^{5/2}  \right]-K_1m_I} \,\,,& \bar{z}_{eq}\leq z< z_{eq}\,,\vspace{0.5cm}\\[6pt]
\dfrac{m_I}{1-\alpha_3\left[ (1+\bar{z}_{eq})^2-(1+z)^2  \right]-\beta_3\left[(1+\bar{z}_{eq})^4-(1+z)^4  \right]-K_2m_I} \,\,,&  z< \bar{z}_{eq}\,.  
\end{array}
\right.
\end{equation}
The constant of integration $m_I$ represents the initial mass of the seed, which forms at $z=z_I$. The term ``seed'' is used here to denote a background fluctuation that eventually generates the black hole. The terms $\{\alpha_j\}$, $\{\beta_j\}$ and $\{K_j\}$ are constants which depend on the Hubble constant, the density parameters today and the mean value of the velocity of dark-matter particles far from the black hole. These quantities are explicitly presented in table~\ref{tabla1}. The values used for the density parameters and the Hubble constant are presented in table~\ref{tabla2}.

\begin{table}[!hbt]
\begin{center}
\renewcommand{\arraystretch}{2.5}
\caption{Constants $\{\alpha_j\}$, $\{\beta_j\}$, $\{K_j\}$ in the mass function $m(z)$, according to Eq.~\eqref{masa}.}
\begin{tabular}{|c|c|c|c|}
\hline
 $\,\,j\,\,$ & $\alpha_j/m_I$ & $\beta_j/m_I$ & $K_jm_I$\\
\hline
 1 &\,\, $\dfrac{6GH_0\Omega_{m0}}{\sqrt{\Omega_{r0}}c^2v_{\infty 0}}$\,\, &\,\, $\dfrac{81GH_0\sqrt{\Omega_{r0}}}{16c^3}$ \,\,&\,\,$\alpha_1\ln\left|\dfrac{1+z_I}{1+z_{eq}}\right|+\beta_1\left[\left(1+z_I\right)^2-\left(1+z_{eq}\right)^2  \right]$\,\,  \\[0.2cm] \hline
2 &\,\, $\dfrac{12GH_0\sqrt{\Omega_{m0}}}{c^2v_{\infty 0}}$ \,\, &\,\, $\dfrac{81GH_0\Omega_{r0}}{20\sqrt{\Omega_{m0}}c^3}$ \,\, &\, $\alpha_2\left[\left(1+z_{eq}\right)^{1/2} - \left(1+\bar{z}_{eq}\right)^{1/2}  \right] +\beta_2\left[\left(1+z_{eq}\right)^{5/2} -\left(1+\bar{z}_{eq}\right)^{5/2}  \right]+K_1m_I$ \,\\[0.2cm] \hline
3 & \,\,$\dfrac{3GH_0\Omega_{m0}}{\sqrt{\Omega_{\Lambda 0}}c^2v_{\infty 0}}$\,\, &\,\,$\dfrac{81GH_0\Omega_{r0}}{32\sqrt{\Omega_{\Lambda 0}}c^3}$\,\, &0\\[0.2cm]
\hline
\end{tabular}\label{tabla1}
\end{center}
\end{table}

\begin{table}[!hbt]
\begin{center}
\renewcommand{\arraystretch}{2.5}
\caption{Assumed values for the density parameters and Hubble constant.}
\begin{tabular}{|c|c|c|c|}
\hline
 $H_0$ & \,\,$\Omega_{m0}$\,\, &\,\, $\Omega_{\Lambda_0}$\,\, & $\,\,\Omega_{r0}\,\,$\\
\hline
$72km\,s^{-1}Mpc^{-1}$ &$0.3$ &$0.7$&$\dfrac{0.415}{H_0^2} km^2s^{-2}Mpc^{-2} $  \\
\hline
\end{tabular}\label{tabla2}
\end{center}
\end{table} 

As it will be seen in next section, the seed formation does not necessarily coincide with the emergence of the apparent horizons defining the black hole.
With the Hubble function and mass function given by Eqs.~\eqref{hubblefunction}-\eqref{masa}, the conditions~\eqref{condi1} to \eqref{m0h0} are satisfied. In particular, requirement~\eqref{m0h0} depends on the time of the formation of the seed $t_{I}$ and the initial mass of the black hole $m_{I}$ because those initial conditions determine the value for the asymptotic mass $m_0$. As a matter of fact, those parameters can lead to a divergence of the mass function for some $z$. In such cases, the  condition~\eqref{m0h0} is not obeyed. However, in the present work, for  nondivergent values of the mass function at $z=0$ the condition is satisfied.

\begin{figure}[h]
\begin{center}
\includegraphics[width=9.5cm]{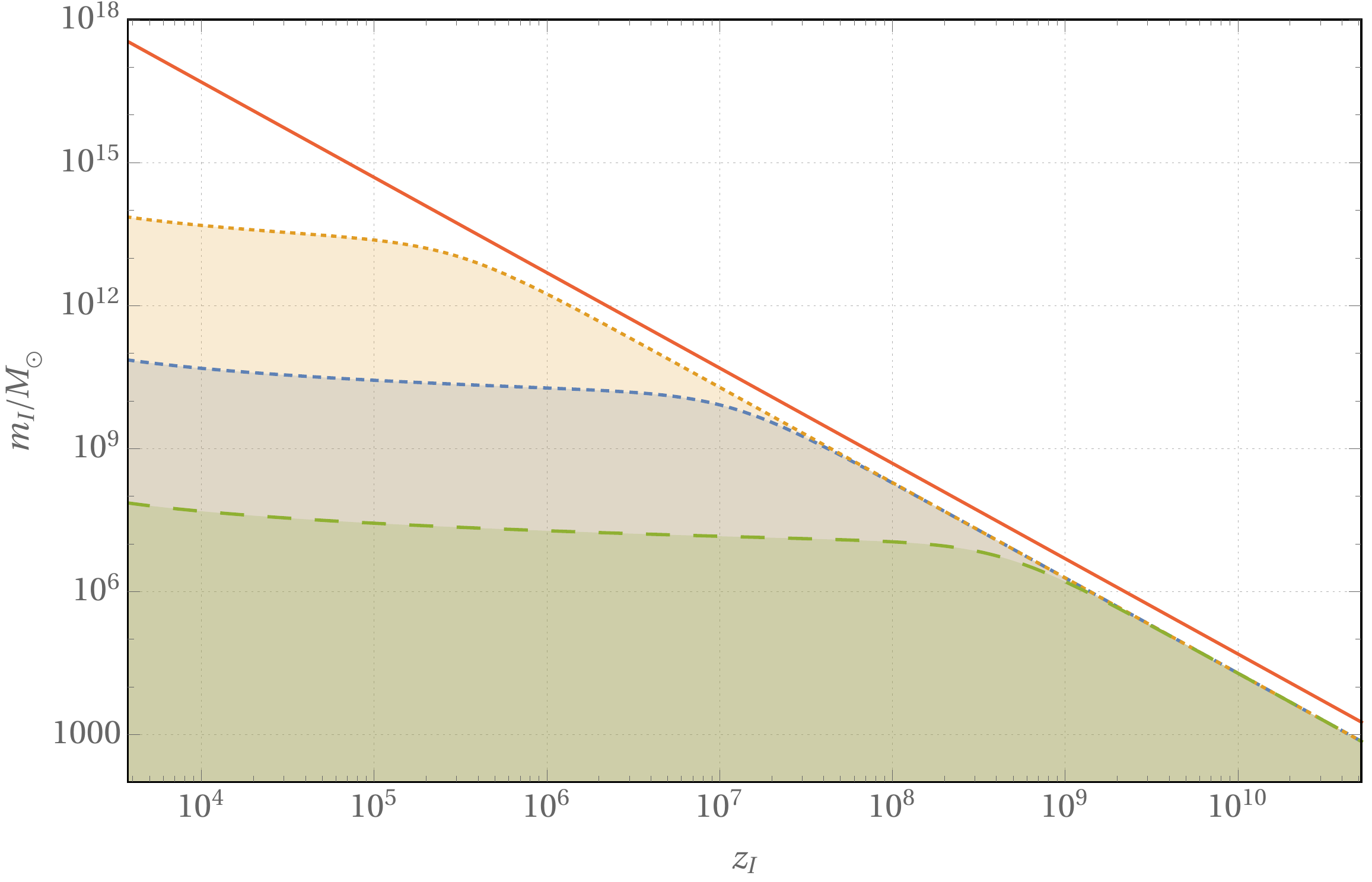}
\caption{
Constraints on $m_I$ and $z_I\,$. The colored region under each curve is the permitted region for the initial conditions $m_I$ and $z_I$ for the black hole in the radiation dominated era with $v_{\infty 0}=10^{-7}pc/yr$ in orange (dotted), $v_{\infty 0}=10^{-10}pc/yr$ in blue (short dashed) and  $v_{\infty 0}=10^{-13}pc/yr$ in green (large dashed). That three curves establish the set of initial conditions that lead to divergence of mass function at $z=0$. The red curve is the particle-horizon mass as a reference. The results for other values of $v_{\infty 0}$ are qualitatively similar.}
\label{fig1}
\end{center}
\end{figure}

\section{Analyzing the primordial black-hole model}
\label{analizing}

\subsection{Constraining the parameter space}

 Given the derived mass function $m(z)$, one may ask about its regularity since a  divergence will appear when the denominator vanishes, according to Eq.~\eqref{masa}. The existence of three different epochs leads us to consider separately the cases in which such divergence may occur. For a fixed velocity $v_{\infty 0}$,  they establish together a constraint on the possible values of $m_I$ and also on the time  $t_I$ demarking the emergence of the black hole.  Actually,  once the value of $v_{\infty 0}$ is fixed, the analysis of the allowed  values of $m_I$ is immediate.

In figure~\ref{fig1} we illustrate the limits of $m_I$ for the radiation era. All curves except the red one (solid curve) correspond to the pairs $(z_I,m_I)$ generating divergences on the mass function~\eqref{masa} today. Thus, every point under each curve is a valid choice to get finite results. Also in this figure we compare the limits of $m_I$ with the particle-horizon mass $m_h$. The quantity $m_h$ denotes the mass within the particle horizon of the FLRW metric surrounding the black hole, being given by
\begin{equation}
m_h =\frac{c^3t}{G} = \frac{c^3}{2GH_0\sqrt{\Omega_{r0}}(1+z)^2 } \, .
\label{result-mh}
\end{equation}
For each time $t$, $m_h$ provides an upper limit for the black-hole mass.
It should be noticed that $m_h$ is greater than the limit for $m_I$. Consequently, there are no black holes with masses of the order of the particle cosmological horizon mass. This result is compatible with a previous analysis made by Hawking and Carr based on different arguments \cite{10.1093/mnras/168.2.399}.  The curve for the divergence can be scaled when other velocities of dark-matter particles are chosen, but the particle-horizon mass curve is not crossed regardless of the value of $v_{\infty 0}$ adopted. The higher the velocity, the smaller the accretion rate, and consequently, the greater the constraint curves will be. The overall picture proposed here presents some similarity with the scenario where intermediate-mass PBHs are the seeds for the supermassive black holes in galactic nuclei \cite{carr-silk}. However, the velocity of dark-matter particles $v_{\infty 0}$ is not easy to estimate. 

It is possible to rethink the question concerning the evolution of the black-hole mass if we rely on the fact that there is no evidence yet for black holes greater than $10^{10}M_{\odot}$ today and  demand a positive velocity for dark-matter particles. By using the mass function $m(z)$ in Eq.~\eqref{masa}, we obtain an implicit relation for the velocity $v_{\infty 0}$. This expression is written in terms of the initial mass $m_{I}$, the current mass value $m(z=0)$ and the redshift of seed creation, of matter-radiation equivalence and radiation-dark energy equivalence $z_I$, $z_{eq}$, and $\bar{z}_{eq}$, respectively, as
\begin{gather}
 \alpha_3\left[ (1+\bar{z}_{eq})^2-1 \right]+\alpha_2\left[ (1+z_{eq})^{1/2}-(1+\bar{z}_{eq})^{1/2} \right] +\alpha_1\ln\left|\dfrac{1+z_I}{1+z_{eq}}\right|  \nonumber \\
= 1-\dfrac{m_I}{m(0)}-\beta_3\left[ (1+\bar{z}_{eq})^4-1 \right] -\beta_2\left[ (1+z_{eq})^{5/2}-(1+\bar{z}_{eq})^{5/2} \right]-\beta_1\left[(1+z_I)^2-(1+z_{eq})^2\right]  \,.
\label{vinf}
\end{gather}  
The implicit dependence of $\{\alpha_j \}$ on $v_{\infty 0}$ allows us to obtain the velocity in terms of the other parameters.

\begin{figure}[h]
\begin{center}
\includegraphics[width=9.5cm]{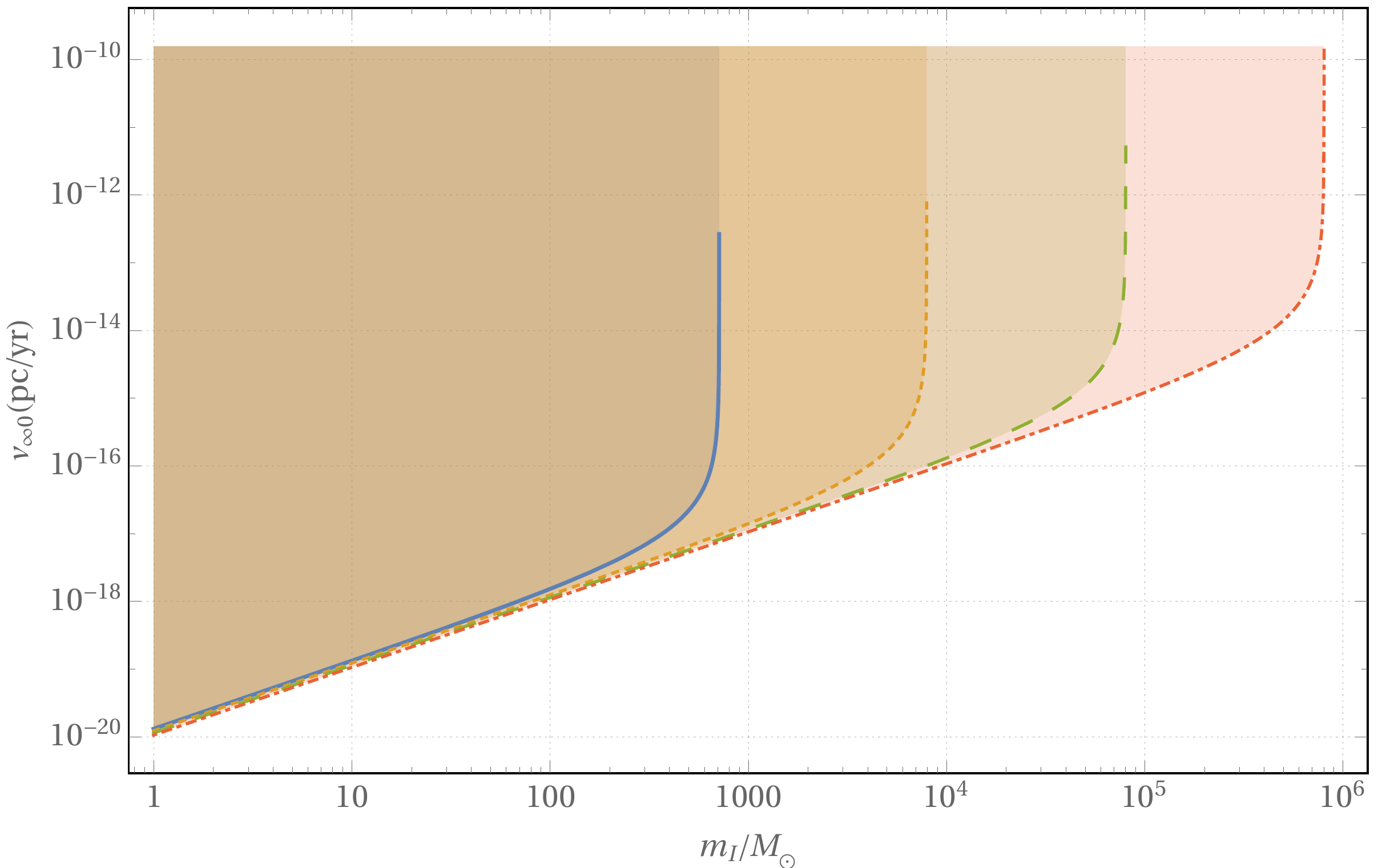}
\caption{Velocity $v_{\infty 0}$ as a function of the mass $m_I\,$. For four different initial times, namely $10^{-2}s$ (blue/thick), $10^{-1}s$ (orange/dotted), $10^0s$ (green/dashed) and $10^1s$ (red/dot-dashed), a mass $10^{10}M_{\odot}$ is produced at $z=0$. The region under each curve generates greater masses than $10^{10}M_{\odot}$ today, whereas the colored region produces the whole range of masses below.} 
\label{fig2}
\end{center}
\end{figure}

Figure~\ref{fig2} shows the dependence of the velocity $v_{\infty 0}$ as a function of  $m_I$ for different initial times. Note that each curve diverges when the initial mass is big enough with any velocity producing masses greater than $10^{10}M_{\odot}$ today. In the next section, a pair $(m_I,v_{\infty 0})$ will be taken for modeling the horizons for which the restrictions  derived here must be considered. Note also that since the velocity $v_{\infty 0}$ must be a positive number, the denominator of Eq.~\eqref{vinf} provides an upper limit for the possible initial seed masses $m_I$. Unlike the restriction posed by the positivity of the mass function (see figure~\ref{fig1}), the velocity does not need to be specified in this case, up to a constraint. More concretely,
\begin{gather}
m_I <
m(0) \left\{
1 + m(0)\left(\dfrac{\beta_3}{m_I}\right)\left[ (1+\bar{z}_{eq})^4-1 \right]
+ m(0)\left(\dfrac{\beta_2}{m_I}\right)\left[ (1+z_{eq})^{5/2}-(1+\bar{z}_{eq})^{5/2} \right] \right.+ \nonumber \\
\left. + m(0)\left(\dfrac{\beta_1}{m_I}\right)\left[ (1+z_I)^2-(1+z_{eq})^2 \right] \right\}^{-1} \, ,
\end{gather}
where the factors $\beta_j/m_I$ only depend on fundamental constants and cosmological parameters, as can be seen in table~\ref{tabla1}. Chosing $m(0)=10^{10}M_{\odot}$ we have an acceptable relation $m_I(z_I)\,$.

Furthermore, we assume that any PBH was originated by an overdensity in the cosmic fluid. The evolution of such a region was found, as well as the initial  size of the overdensity \cite{carr-spectrum}. In the present work, the overdensity is considered at the time of the black-hole formation. 
The condition for a spherical region of a fluid to collapse into a black hole is that its radius should be smaller than the Schwarzschild radius, 
$r < r_s \equiv 2Gm/c^2$ and, as such, its volume as seen by an external observer far apart in the cosmological background is nearly $V<4\pi r_s^3/3$, while its density is greater than a critical value 
\begin{equation}
\bar{\rho} = \rho + \delta \rho > \frac{3c^6}{32\pi G^3m_I^2} \, . 
\end{equation}
In previous expression, $\rho$ denotes the background density \cite{10.1093/mnras/168.2.399}. Now, defining the relative density contrast $\delta$ as
\begin{equation}
\delta \equiv \frac{\Delta \rho}{\rho} 
= \frac{\bar{\rho}-\rho}{\rho} \, ,
\end{equation}
the following inequalities should be respected:
\begin{equation}
\delta > \frac{3c^6}{32\pi G^3m_I^2\rho} - 1 > 0 \,.
\end{equation}
The left-hand side inequality takes into account the condition for the formation and the right-hand side guarantees that the density contrast is in fact an overdensity. Both relations lead to 
\begin{eqnarray}
m_I&<\dfrac{c^3}{2GH_0\sqrt{\Omega_{r0}}(1+z_I)^2}=m_h(z_I)\,,\label{mi1}\\
m_I&\!\!\!\!\!\!\!\!\!>\dfrac{c^3}{2GH_0\sqrt{\Omega_{r0}}\sqrt{1+\delta}(1+z)^2}\, ,
\label{mi2}
\end{eqnarray}
which means that black hole with the cosmological horizon mass formed through overdensities are forbidden. The previous inequalities set a region in the $(m_I,z_I)$ space in which the black-hole formation is possible.

\begin{figure}[h]
\begin{center}
\includegraphics[width=9.5cm]{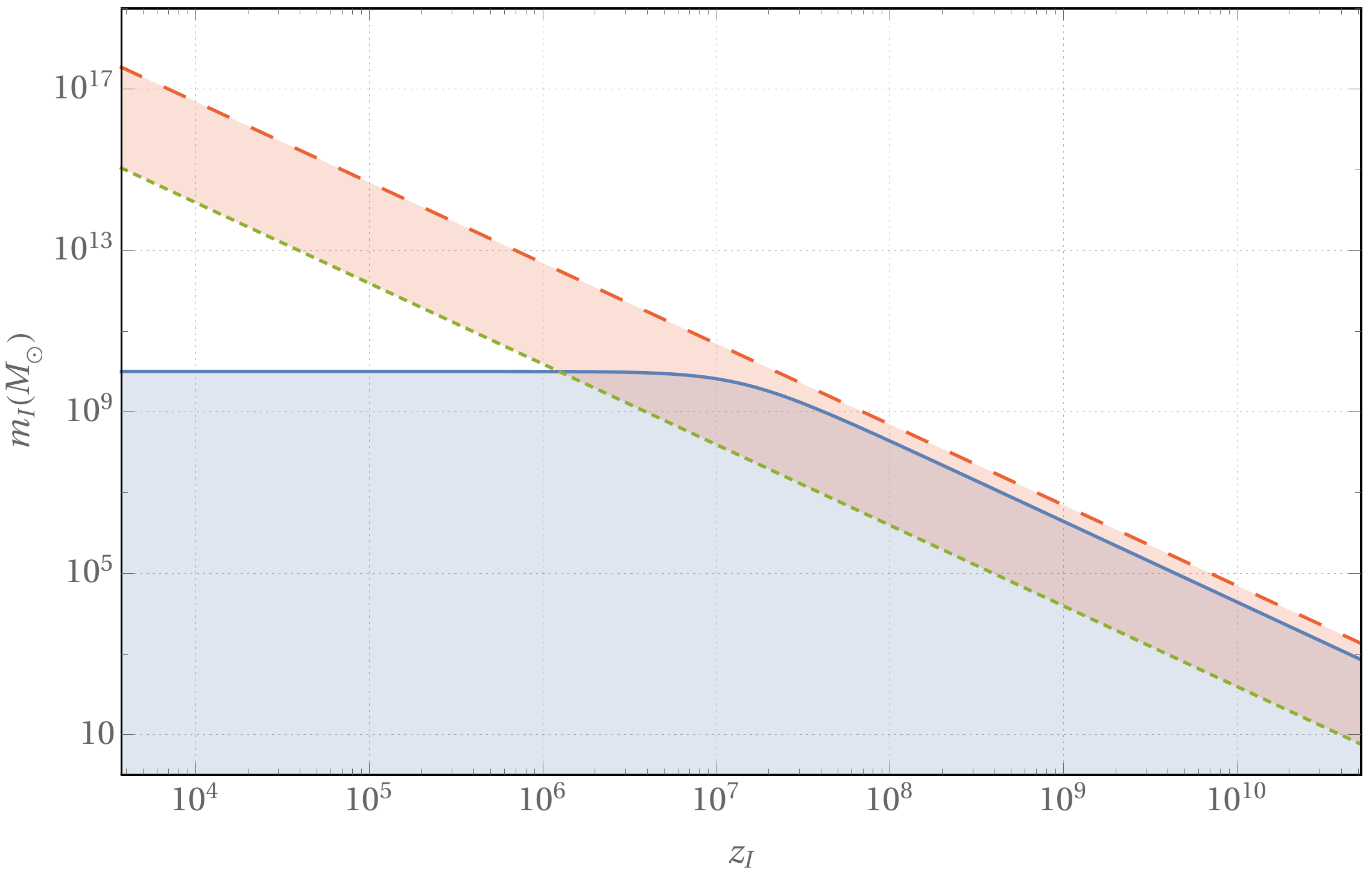}
\caption{Formation and finitude constraint. The straight stripe in red indicates the black-hole formation having the upper limit the particle-horizon mass. The blue region under the thick curve indicates initial conditions that generate black holes compatible with the observations today. The overlapping of them represents initial conditions allowing the PBH formation and its growth to $10^{10}M_{\odot}$ (at most), simultaneously. In this plot, the $\delta=10^5$ has been assumed.}
\label{fig3}
\end{center}
\end{figure}

In figure~\ref{fig3} we show the resulting constraint by taking into account the accretion limitations and the PBH formation by overdensities. It should be noted that the region under the thick curve indicates those PBHs accreting mass until $10^{10}M_{\odot}$ today. The region enclosed by the straight lines allows a PBH formation depending on the value of the density contrast at the time of the formation. For $\delta \approx 5$ and up the overlapping blue-red region allowing PBHs with masses under $10^{10}M_{\odot}$ today is not an empty set.

\subsection{Evolution of the horizons}

Let us now discuss the expected horizon evolution in the framework of the present model. Apparent horizons can be represented as curves in the $(\hat{r},z)$ space. To obtain the solutions is necessary to fix the redshift for the black-hole creation $z_I$ (or the initial time $t_I$), the initial mass $m_I$, and the velocity of the accreted particles $v_{\infty 0}$ far from the black hole. Choosing $t_I=10^{-2}s$ 
($z_I \approx 5\times 10^{10}$) as the time for the initial seed (overdensity fluctuation), the constraints presented in figures~\ref{fig3} and \ref{fig2} allow us to select a mass value in the range $(5.8M_{\odot},713.9M_{\odot})$. 

\begin{figure}[h]
\begin{center}
\includegraphics[width=9.5cm]{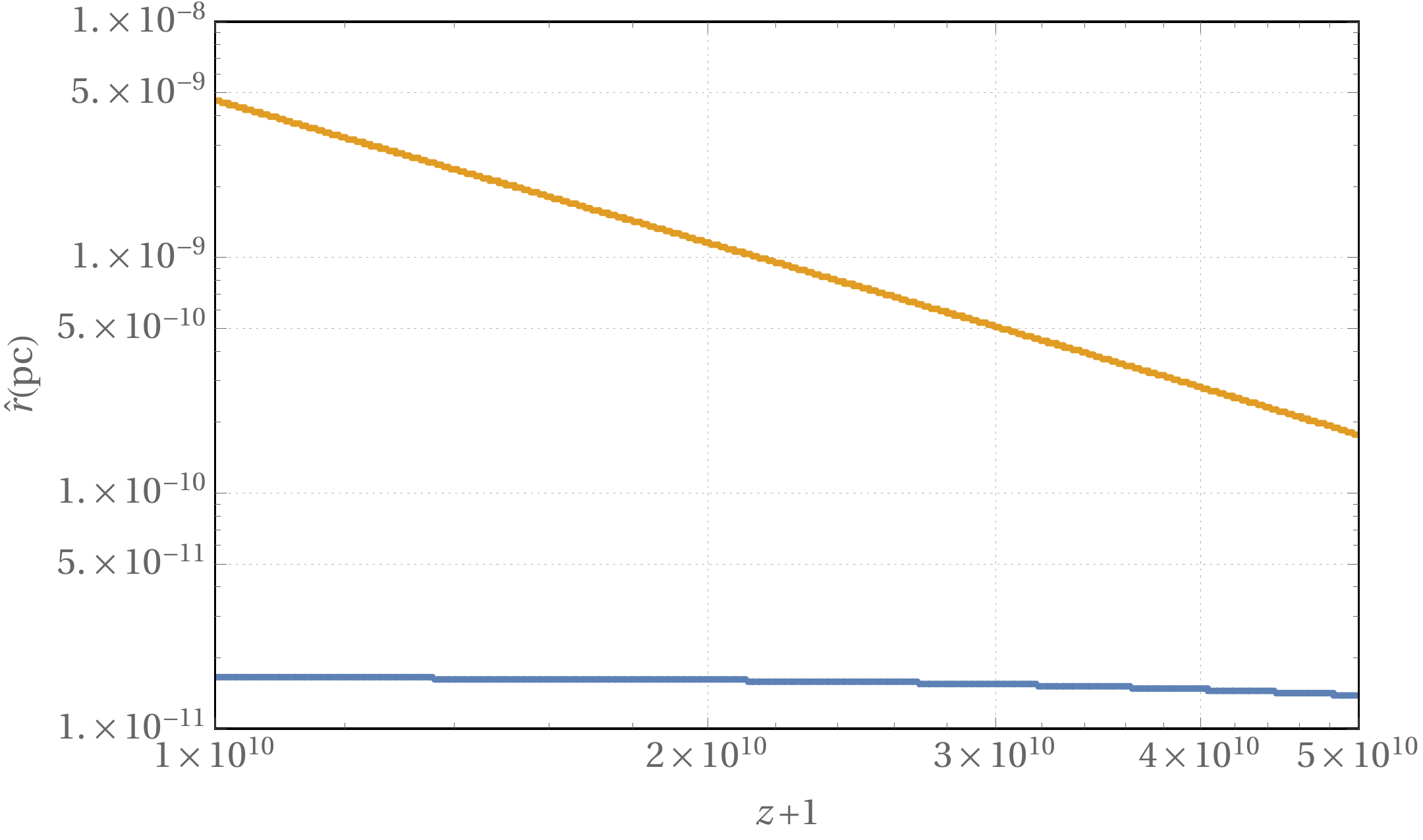}
\caption{Apparent-horizon radius as a function of $(z+1)$.  The cosmological horizon is the upper curve in orange, while the lower curve in blue is the black-hole horizon. For this graph, $m_I=140M_{\odot}$, $t_I=10^{-2}s\,(z_I\approx 5.2\times10^{10} )$, $v_{\infty 0}=10^{-10}pc/yr$.}  
\label{fig4}
\end{center}
\end{figure}

In figure~\ref{fig4}, by assuming $m_I=140M_{\odot}\ll m_h(z_I)\approx 1800M_{\odot}$ and velocity  $v_{\infty 0}=10^{-10}pc/yr$, the typical horizon structure is displayed. 
The plot starts with the creation of the seed time. There are two marginal surfaces corresponding to the two horizons. Note that the black-hole horizon radius (first curve from bottom to top) is smaller than the cosmological horizon radius and it evolves at a similar rate to that of the singularity $\hat{r}=2Gm/c^2$.

\begin{figure}[h]
\begin{center}
\includegraphics[width=9.5cm]{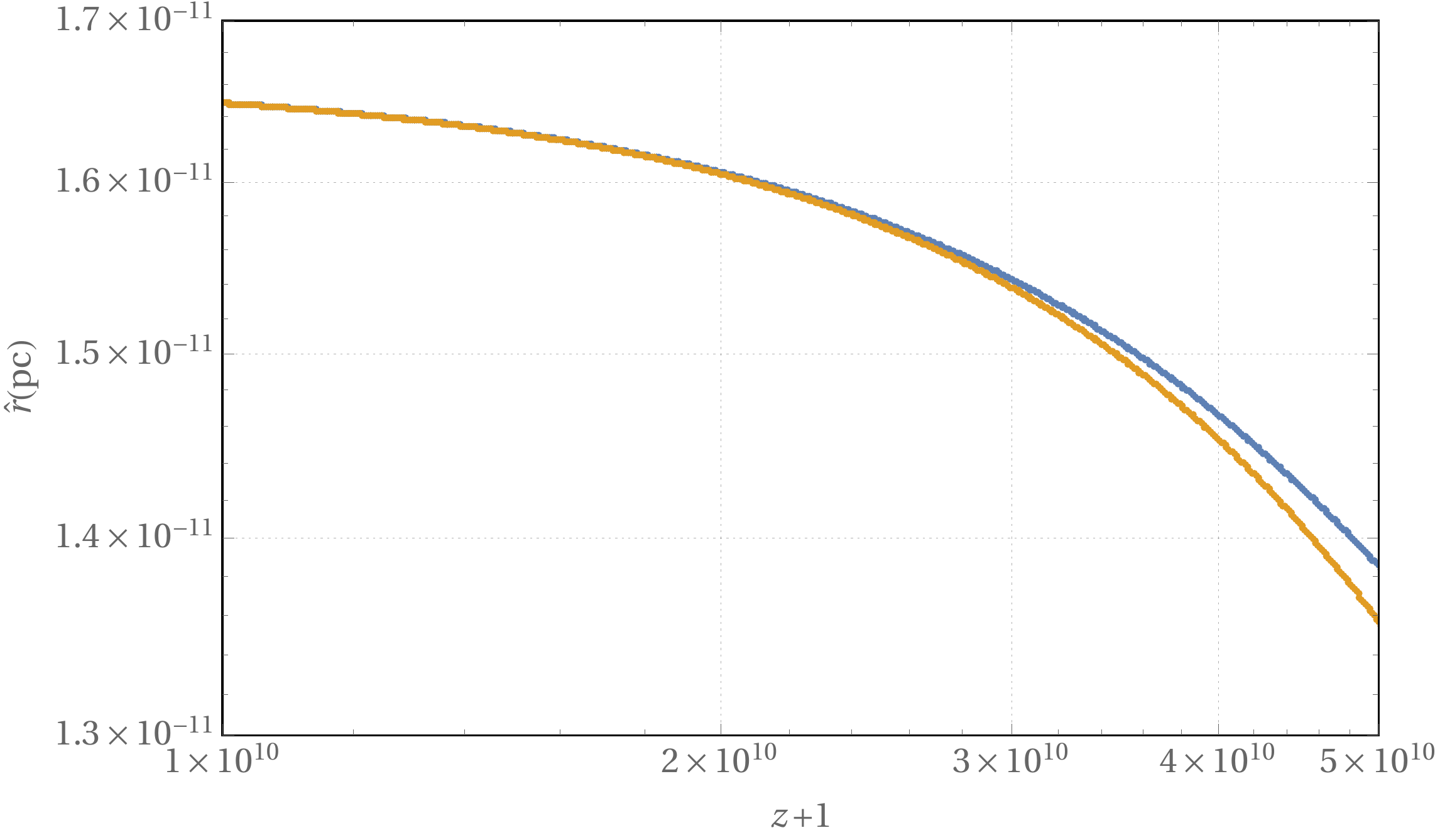}
\caption{A more detailed view of the black-hole horizon structure of the figure~\ref{fig4} in blue (upper curve) and the singularity as a reference in orange (lower curve). Both curves have the same asymptotic behavior at low redshifts.}  
\label{fig5}
\end{center}
\end{figure}

In figure~\ref{fig5}, the singularity (first curve from bottom to top) and the horizon have the same asymptotic behavior as the accretion rate decreases, but the horizon is always over the singularity. The expansion of null inner vector fields orthogonal to the spheres $(t_0,r_0,\theta,\phi)$ is null in the apparent horizon, and different from zero into the inner spheres containing the singularity.

\begin{figure}[h]
\begin{center}
\includegraphics[width=9.5cm]{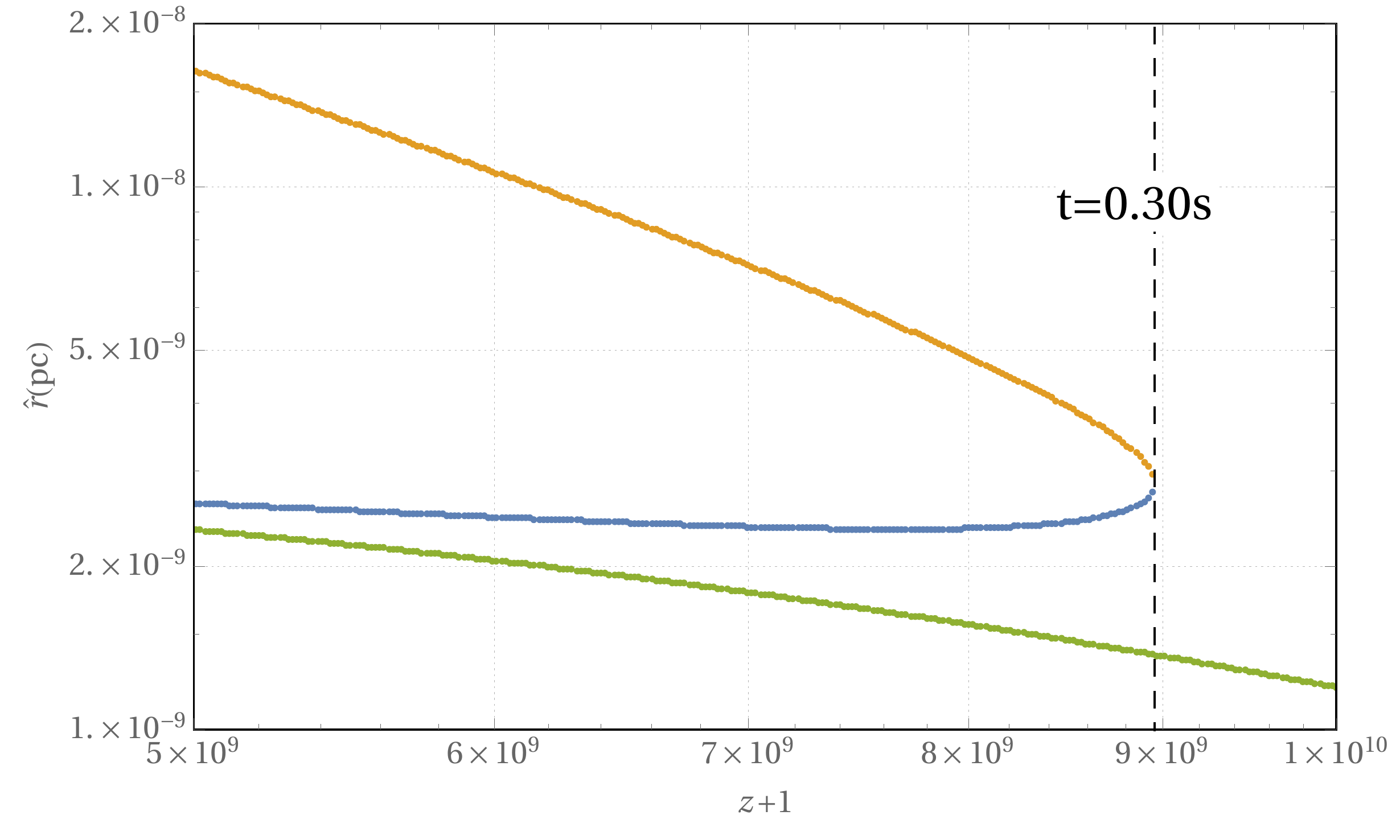}
\caption{From the bottom curve to the upper curve: the singularity as a reference (green), the black-hole horizon (blue), and the cosmological horizon (orange). The horizons originate at $t\approx0.30s$. For this graph, $m_I=700M_{\odot}$, $t_I=10^{-2}s\,(z_I\approx 5.2\times10^{10} )$, $v_{\infty 0}=10^{-10}pc/yr$. The singularity approaches the black-hole horizon asymptotically as $z$ decreases ($t$ grows).} 
\label{fig6}
\end{center}
\end{figure}

A different qualitative behavior appears when the initial mass approaches the particle cosmological horizon mass at the time of formation. As an illustration, we consider $m_I=700M_{\odot}\lessapprox m_h(z_I)$ and $v_{\infty 0}=10^{-10}pc/yr$. These parameters are indicated in the permitted region of figure~\ref{fig2}. The solutions are shown in figure~\ref{fig6}, and for completeness the corresponding singularity curve has been added. One important point to be noticed is that the time in which the real and non-negative solutions emerge at $t=0.3s$ is different from the time of the black-hole formation, $t_i=10^{-2}s$. Such a delay defines a kind of incubation time, $t_{inc}$. Initially, the two spacelike regions beyond the corresponding horizons are joined before $t \approx 0.30s$. 

Actually, as the expansion proceeds and overcomes the accretion on the black hole, the timelike region appears. Thus the incubation time is the necessary time to produce a timelike region in the causal structure.%
\footnote{In the theory of galaxy formation, there is also an incubation time, $\tau_{inc}$. It is defined as an estimate of the amount of time interval from the beginning of structure formation process in the universe until the formation time \cite{Sand1993,Lima:2007yz}. Here it is related with the time needed to the formation of a primordial black hole.}
The condition to obtain a non-null incubation time is roughly that the particle-horizon mass $m_h$ in the pure FLRW cosmology at $z_I$ is close to $4m_I$ and its magnitude depends on the initial mass. The black-hole influence on the spacetime in the vicinity of the cosmological horizon is to pull it, so the cosmological horizon is slightly modified when compared with the pure FLRW case at times of the order of $t_I + t_{inc}$.

\begin{figure}[h]
\begin{center}
\includegraphics[width=9.5cm]{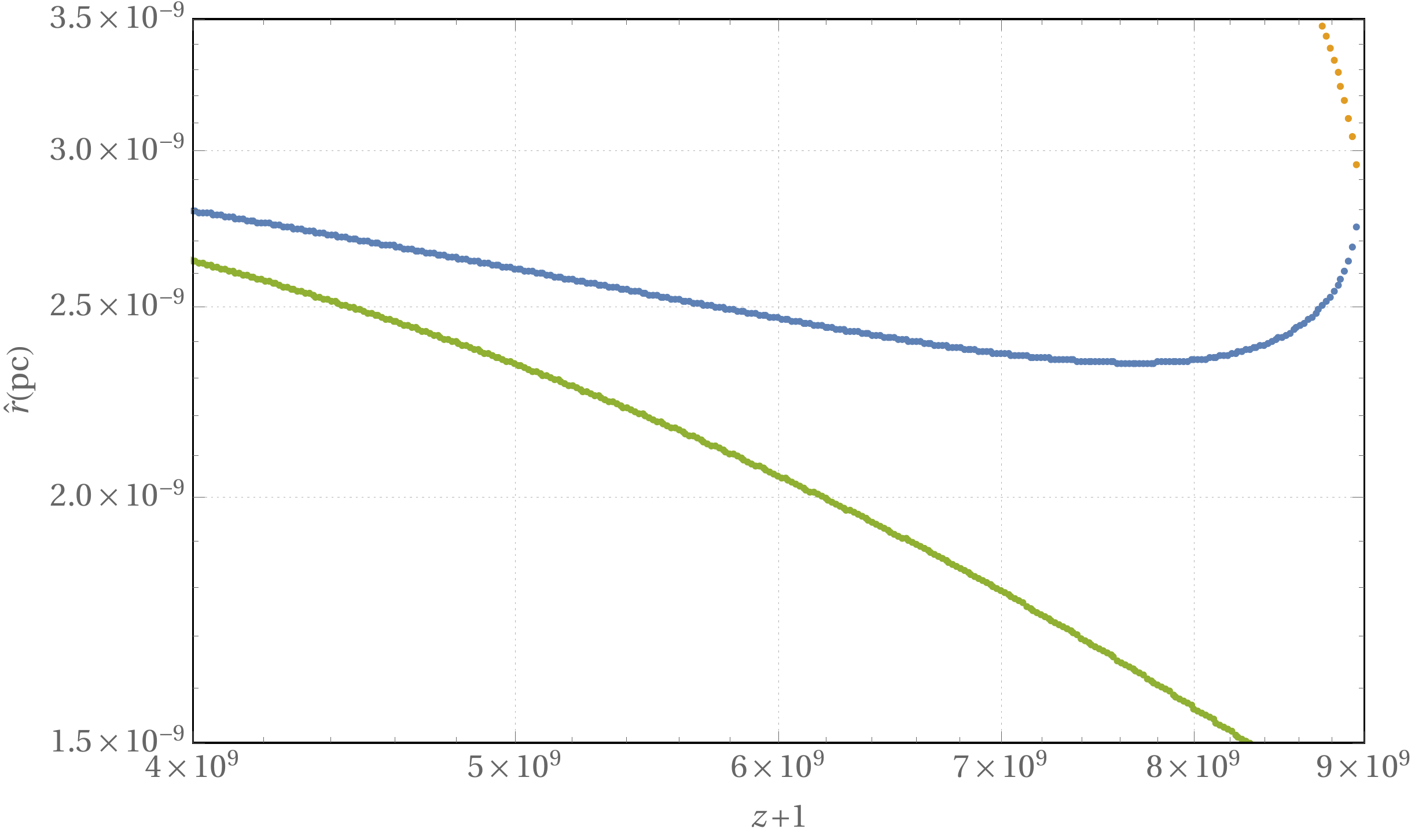}
\caption{From the bottom curve to the upper curve: the singularity (green), the black-hole horizon (blue), and the cosmological horizon (orange). The influence of the cosmological horizon on the local dynamics near to the black hole is noted by the distancing between the singularity and the horizon for high redshift. For this graph, $m_I=700M_{\odot}$, $t_I=10^{-2}s\,(z_I\approx 5.2\times10^{10} )$, $v_{\infty 0}=10^{-10}pc/yr$.}
\label{fig7}
\end{center}
\end{figure}

In figure~\ref{fig7}(a) a more detailed view of the black-hole horizon and the singularity is presented. The interaction between the horizons produces the unexpected local decreasing in size of the black-hole horizon and its separation from the singularity.

\begin{figure}[h]
\begin{center}
\includegraphics[width=9.5cm]{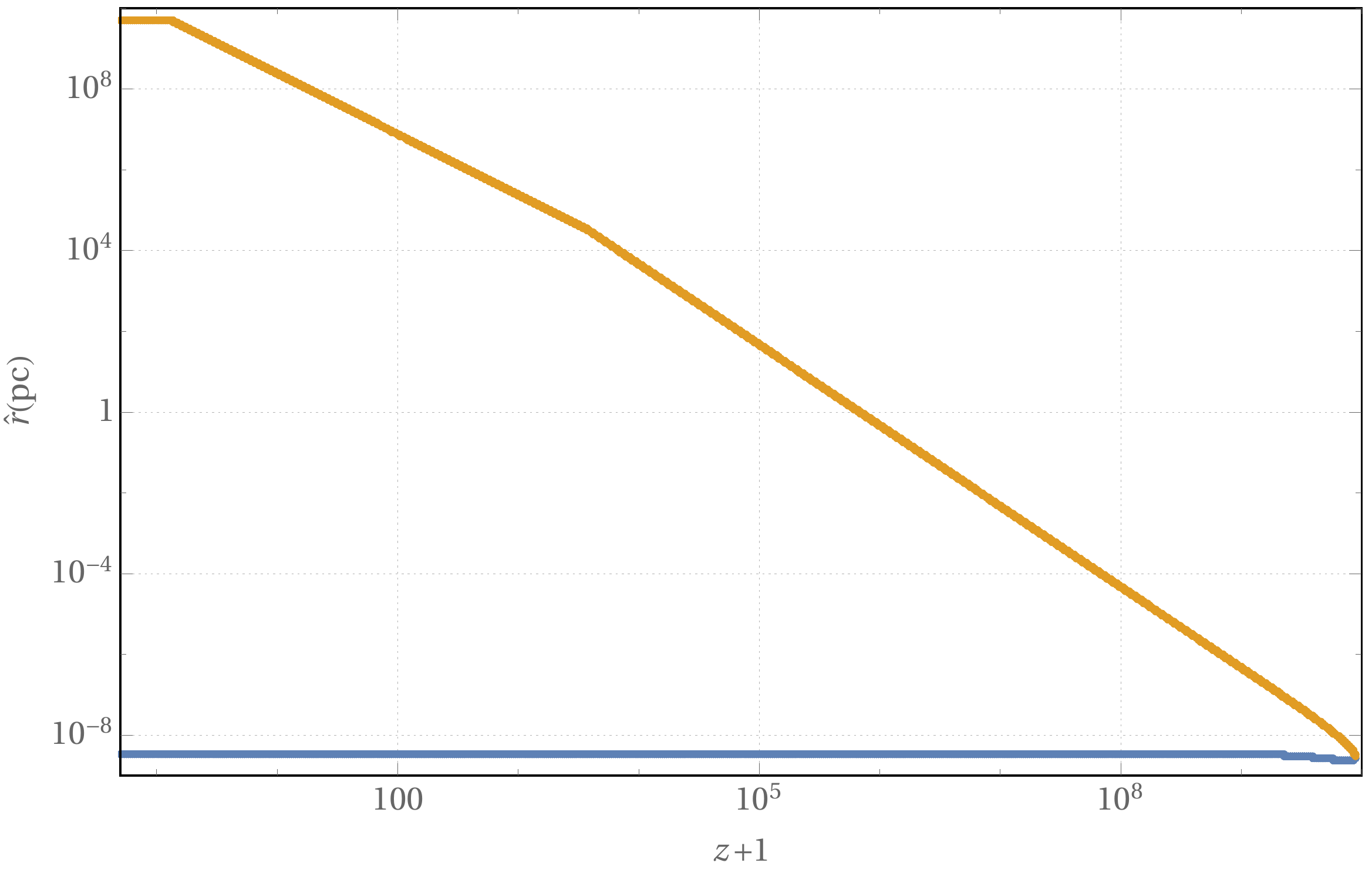}
\caption{The large scale horizon structure, the black-hole horizon (blue curve) at bottom and cosmological horizon (orange curve) is the upper curve. At large scales, the dynamical black hole looks approximately as a Schwarzschild black hole. For this graph, $m_I=700M_{\odot}$, $t_I=10^{-2}s\,(z_I\approx 5.2\times10^{10} )$, $v_{\infty 0}=10^{-10}pc/yr$.} 
\label{fig8}
\end{center}
\end{figure}

The dynamical behavior of the horizons can be analyzed throughout the whole interval $z\leq z_I$ but it is remarkably stronger for higher redshifts. In fact, at cosmological timescales, the two solutions look as shown in figure~\ref{fig8}. The black-hole horizon is approximately constant though the accretion mechanism does not stop ever, while the cosmological horizon is a piecewise function, reflecting the fact that the Hubble function was also piecewise defined.

The cosmological horizon, the black-hole horizon, and black-hole mass today are 
\begin{equation}
\hat{r}_h=4.98 Gpc\,, 
\,\, \hat{r}_{bh}=1.0564\times 10^5 km\,,
\,\, m_{bh}=35850.8M_{\odot}\,.
\label{num-values}
\end{equation}
The cosmological horizon radius presented in~\eqref{num-values} is compatible with the accepted value for the size of the observable Universe today. In fact, considering typical values for the initial mass of the seed, the large-scale structure is not affected by the local physics. If the mass of the black hole today is $10^{10}M_{\odot}$, the cosmological horizon does not suffer any appreciable change, maintaining its value $\hat{r}_h$. This implies that there is no way to detect the influence of the super massive black holes known today by means of the behavior of the cosmological horizon.

Keeping in mind the several choices of initial times of black-hole creation,  the same qualitative behavior as shown in figures~\ref{fig4} to \ref{fig8} is observed. One interesting feature of the  accretion dynamics discussed here is that the photon accretion in the radiation era is considerably relevant. Higher $z_I$ implies that the black hole can grow more efficiently during the radiation era. In this way, the PBHs can be big enough in the matter-dominated era thereby reaching the intermediate and higher masses now observed. The impact of the velocity parameter $v_{\infty 0}$ is also important. It is more difficult for the black hole to accrete matter for higher initial particle velocities.

\section{Final remarks}
\label{final-remarks}

Formation and evolution of primordial black holes are dynamically discussed based on a class of  McVittie-type solutions. Such models are compatible with the Schwarzschild and FLRW geometries thereby interpolating the extremely local and large-scale cosmological structures. Dark matter and radiation are considered as the primary components contributing for the accretion process. The former is described as a pressureless fluid of noninteracting particles.

In the present work, it is assumed that the origin of the primordial black hole is an overdensity in the primeval cosmic fluid, considering constraints associated with the initial perturbation. The relevant parameters of our model are the mass of the initial perturbation at the moment of collapse, the time of formation and the velocity of dark-matter particles away from the central core. Limits on those parameters were derived. Two different scenarios are treated, with small and large mass seed compared to the particle cosmological horizon mass. An incubation time for the emergence of the horizons is identified when the initial mass of the seed is close to the particle-horizon mass.

It is observed that the evolution of the apparent horizons in McVittie-type spacetimes plays an important role in regimes where the dynamics cannot be neglected thereby resulting a complex causal structure. Nevertheless, the geometry can be well described by the Schwarzschild-de Sitter solution in regions where dynamical processes are negligible. Furthermore, the expansion of the Universe at high redshifts can be influenced by the evolution of the black hole, although being fully neglected today. The Schwarzschild description for the black hole is justified locally since its horizon structure is very similar to the Schwarzschild case.  

We conclude that, by taking into account what is currently known about the primordial universe phenomenology, generalized McVittie spacetimes are plausible candidates for describing primordial black holes. A more complete and realistic scenario it will be discussed in a forthcoming communication.

\newpage

\begin{acknowledgments}

F.R. acknowledges the support of National Council for Scientific and Technological Development (CNPq), Brazil. 
C.M. is supported by Grant No 420878/2016-5, National Council for Scientific and Technological Development (CNPq), Brazil. 
J.A.S.L. is partially supported by
National Council for Scientific and Technological Development (CNPq), Grant No 310038/2019-7; Coordination for the Improvement of Higher Education Personnel (CAPES), Grant No 88881.068485/2014; and S\~ao Paulo Research Foundation (FAPESP), LLAMA Project No 11/51676-9.

\end{acknowledgments}

\end{document}